\documentclass[aps,prb,twocolumn,showpacs,preprintnumbers,amsmath,amssymb]{revtex4}
\usepackage{dcolumn}
\usepackage{bm}
\usepackage{amsmath}
\usepackage{feynmp}

\usepackage{graphicx}
\begin{document}
\title{Quantized Repetitions of the Cuprate Pseudogap Line}
\author{Vincent Sacksteder IV}
\email{vincent@sacksteder.com}
\affiliation{Department of Physics, Royal Holloway University of London, Egham Hill, Egham, TW20 0EX, United Kingdom}


\begin{abstract}
 The cuprate superconductors display 
 several characteristic temperatures which decrease as the material composition is doped,
 tracing lines across the temperature-doping phase diagram.  Foremost among these is the pseudogap transition.
 At a higher temperature  a peak is seen in the magnetic susceptibility, and changes in symmetry and in transport are seen at other characteristic temperatures. We report a meta-analysis of all measurements of characteristic temperatures well above $T_c$ in strontium doped lanthanum cuprate (LSCO) and oxygen doped YBCO.   The experimental corpus shows that the pseudogap line is one of a family of four straight lines  which stretches across the phase diagram from low to high doping, and from $T_c$ up to $700$ K.  These lines all  originate from a single point near the overdoped limit of the superconducting phase and increase as doping is reduced. The  slope of the pseudogap lines is quantized, with the second, third, and fourth  lines having slopes that are respectively $1/2,\;1/3,$ and $1/4$ of the slope of the highest  line.      This pattern  suggests that the cuprates host a single mother phase  controlled by a  2-D sheet density which is largest at zero doping and which decreases linearly with hole density, and that the pseudogap lines, charge density wave order, and superconductivity are all subsidiary effects supported by the mother phase.

\end{abstract}

\maketitle

The cuprate high $T_c$ superconductors are built from copper oxide planes, and when these planes are doped with holes superconductivity occurs. Temperatures above the superconducting critical temperature $T_c$ make the cuprates effectively two dimensional by disrupting the weak interplane  coupling. 
 Nonetheless the cuprates  exhibit a wide variety of mysterious features at temperatures far exceeding $T_c$.  One of the first to be noticed is a broad peak in the magnetic susceptibility at a temperature $T_{max}$, which is accompanied by qualitative changes in electrical and thermal transport.  \cite{PhysRevLett.62.957,yoshizaki1990magnetic} At low dopings $T_{max}$ reaches temperatures as high as $700$ K, and it  decreases linearly as the parent compound is doped with holes.  Focus has since shifted toward a pseudogap temperature $T^*$ which is is substantially smaller than $T_{max}$ and again decreases linearly with hole doping. $T^*$  marks a depletion in the density of states, and has been measured by many experimental techniques including angularly resolved photoemission spectroscopy (ARPES).  \cite{norman2005pseudogap,vishik2018photoemission} Over the years a perplexing cornucopia of other doping-dependent signals have been identified, including a lower pseudogap temperature $T_\nu$, charge density waves, nematic order, and a linear resistance extending to very high temperatures marking the "strange metal" regime.  \cite{RevModPhys.87.457,keimer2015quantum}  



In this paper we identify an unexpected regularity in the cuprate phase diagram: four of the experimentally measured doping-dependent temperatures, including $T_{max}$,
 the pseudogap temperature $T^*$, and  the lower pseudogap line $T_\nu$, belong to a family which radiates from a common intersection and has slopes  determined by a quantization rule.  Our results are based on a comprehensive and exhaustive survey of  research papers which measure temperature dependence and report numerical values of characteristic temperatures well above $T_c$.  We report  only  characteristic temperatures at which  clearly identifiable signals occur, for example  peaks or kinks in the temperature dependence, or extinction of a diffraction peak at a particular temperature.   In particular, the reported temperatures have been realized experimentally, rather than being derived by extrapolation from lower temperatures. 

  In order to remove any confusion about material-specific aspects of the phase diagram, we have separated data obtained from different compounds and doping techniques.  We first report in Section \ref{LSCODataText} the characteristic temperatures of strontium-doped La$_{2-x}$Sr$_x$CuO$_4$ (LSCO).  Separately, in Section \ref{YBCODataText} we report the characteristic temperatures of oxygen-doped YBa$_2$Cu$_3$O$_{6+\delta}$ (YBCO). These two compounds and doping strategies, plus a third compound Bi$_2$Sr$_2$Ca$_{n-1}$Cu$_n$O$_{2n+4+\delta}$ (BSCCO),  have been used for the large majority of data on characteristic temperatures of cuprates well above $T_c$; our restriction to LSCO and YBCO is not terribly selective. In the following sections we compare the LSCO and YBCO data and draw conclusions about which aspects  of the high $T_c$ phase diagram are universal and which are material specific. 
  
 
In gathering data we have followed standard practice among  reviews of the high $T_c$ phase diagram and the pseudogap by  focusing on   those experimental signals which are understood to reflect the carriers that mediate superconductivity, and omitting the extensive experimental literature that focuses on ionic behavior.   Along these lines 
 we omit  thermal history dependence and hysteresis, oxygen movement and ordering, and mechanically-oriented observables such as lattice constants and internal friction.

Within these restrictions we have made every effort to be comprehensive.  Our goal has been to  gather into one place all experimental data on characteristic temperatures above $T_c$.

 We  separate the results of distinct experimental groups into distinct data sets.  When a group used several experimental techniques, identified several characteristic temperatures in their data, or re-analyzed data from other publications, we separate these distinct results into distinct data sets.  When a group republished or revised their results, we use the latest results. Full details about each data set and about which papers did and did not match our selection criteria are included in the appendices.  


\section{LSCO data\label{LSCODataText}} 
\begin{figure}[]
\includegraphics[width=9.5cm,clip,angle=0]{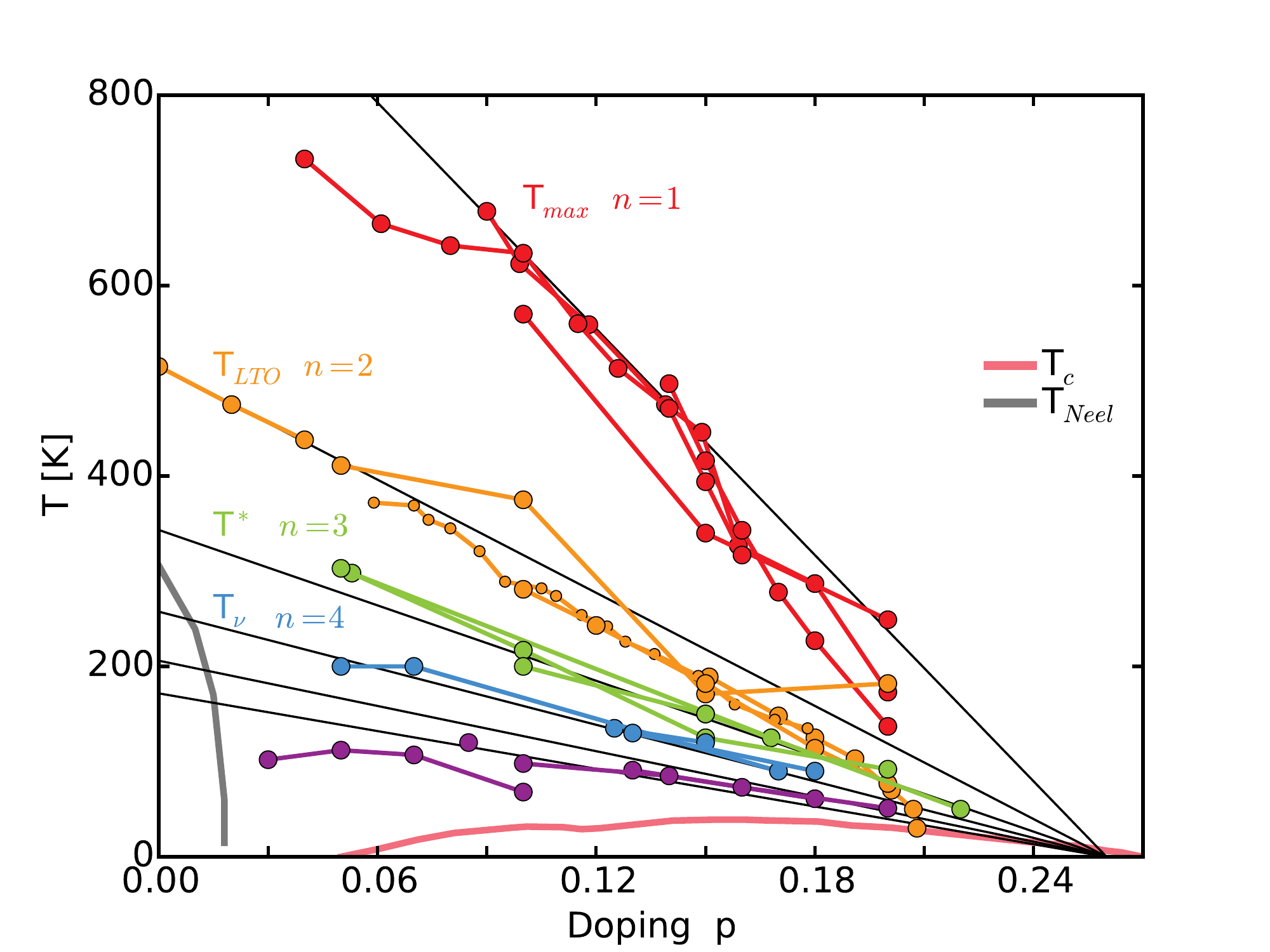}
\caption{ (Color online.) The  family of pseudogap lines in LSCO.  The black guides to the eye have slopes proportional to $1/n$ and meet at doping $p=0.26$.   Individual pseudogap data sets, each from a distinct experimental group measuring a distinct experimental probe and signature, are shown as filled circles connected by lines.  The hallmark of the first (red) pseudogap line $T_{max}$ is a peak in the magnetic susceptibility.  $C_4$ symmetry is broken on the second (orange) line $T_{LTO}$, and the ARPES density of states develops a pseudogap on the third (green) line $T^*$.  The fourth pseudogap line $T_\nu$ (blue) is marked by transport signatures.  Purple lines show NMR, Hall angle, heat capacity, and neutron scattering data sets.  The pink line shows the superconducting $T_c$.  An antiferromagnetic phase (grey line) is found at low dopings.
}
\label{LSCOData}
\end{figure}

Figure \ref{LSCOData} summarizes the experimental corpus on strontium-doped LSCO.  While this compound has a relatively low maximum $T_c$ of about $39$ K, it is stable up to $1000$ Kelvin and  the full range of dopings across the  cuprate superconducting dome can be explored.
The corpus contains a total of twenty distinct data sets from thirteen distinct experimental groups published between 1990 to 2018, using a variety of sample preparation techniques and experimental probes developed over the decades.  Five more data sets \footnote{These sets are labeled in the appendix as $n=3$ set (e) and $n=4$ sets (d,e,f,g).} exist on characteristic temperatures in neodymium and europium-doped LSCO, with excellent agreement with the strontium doped results shown here.   Appendix \ref{LSCOAppendix} gives full details on each  of these data sets and on three more strontium-doped  data sets \footnote{These sets are labeled in the appendix as Omitted Data Sets  (a,p,q).} which we have omitted on a case by case basis. 
In order to show the slope and profile of individual data sets we plot lines connecting the  data points within each set.  

 Figure \ref{LSCOData}  shows that the data sets are in remarkable  agreement, forming a pattern of four distinct characteristic temperatures which each start high at small doping and decrease as doping is increased.   Despite the variety of samples and probes, the scatter in data points around each characteristic temperature is small,  of order $\pm 15$ K on the lowest three characteristic temperatures (orange, green, and blue lines), and somewhat larger on the highest characteristic temperature (red lines).   Gaps  which are much larger than this scatter separate each of the characteristic temperatures, allowing each data set to be unambiguously assigned to one of the  four characteristic temperatures.   We number characteristic temperatures from the top, with $n=1$ on the highest (red) line, $n=2$ on the next highest  (orange), etc.
 
    The topmost red lines (four data sets) show  $T_{max}$, visible both as a peak in the magnetic susceptibility and as kinks in the resistance and thermoelectric power (TEP). \cite{yoshizaki1990magnetic,PhysRevB.49.16000,kim2004two} 
   The yellow lines (second from the top, five data sets) show the temperature $T_{LTO}$ of the well-known transition from the high symmetry tetragonal phase  to a lower symmetry orthorhombic phase, which has generally  though not unanimously \cite{PhysRevLett.68.3777} been regarded as simply a structural phase transition with little relation to high $T_c$.  \cite{kim2004two,PhysRevLett.68.3777,PhysRevLett.93.267001,PhysRevB.57.6165,PhysRevB.46.14034}  This transition manifests clearly in neutron scattering, X-ray diffraction, resistance, and TEP measurements.  
   The green lines (third from the top, four data sets) mark  the pseudogap temperature $T^*$, where  ARPES shows that a pseudogap opens in the density of states.  Like $T_{max}$ and $T_{LTO}$, $T^*$ is  seen also in the resistance and the TEP. \cite{PhysRevLett.103.037004,PhysRevLett.93.267001,kim2004two,PhysRevB.75.140503,PhysRevB.92.134524}.  
The blue lines (fourth from the top, three data sets) are a recently identified second pseudogap transition $T_\nu$ visible in the Nernst effect 
 and in nuclear magnetic resonance (NMR). \cite{FUJII2010S21,PhysRevB.97.064502,xu2000vortex,PhysRevB.64.224519,PhysRevB.69.184503} This transition is well attested by four additional Nernst and resistance data sets \footnote{These sets are labeled in the appendix as  $n=4$ sets (d,e,f,g).} in neodymium-doped and europium-doped variants of LSCO, which give temperatures that agree very well with the strontium-doped LSCO data.   \cite{PhysRevB.95.224517,PhysRevB.97.064502}  We include also four NMR, Hall angle, heat capacity, and neutron scattering data sets  (in purple) which suggest similar lines at lower temperatures. \cite{PhysRevB.69.184503,PhysRevLett.105.027004,xu2000charge,matsuzaki2004electronic}

The fact that over the last thirty years experiments on LSCO's temperature dependence consistently have found distinctive behavior (peaks, kinks, extinction of diffraction  peaks, etc.) on these four fairly crisp lines, and not  in the intervening  gaps, indicates that this pattern has predictive power.  We expect that future experiments will continue to find smooth temperature dependence in the gaps and anomalous behaviors on these lines.  This observed pattern and prediction are our most important results.

       \begin{figure}[]
\includegraphics[width=9.5cm,clip,angle=0]{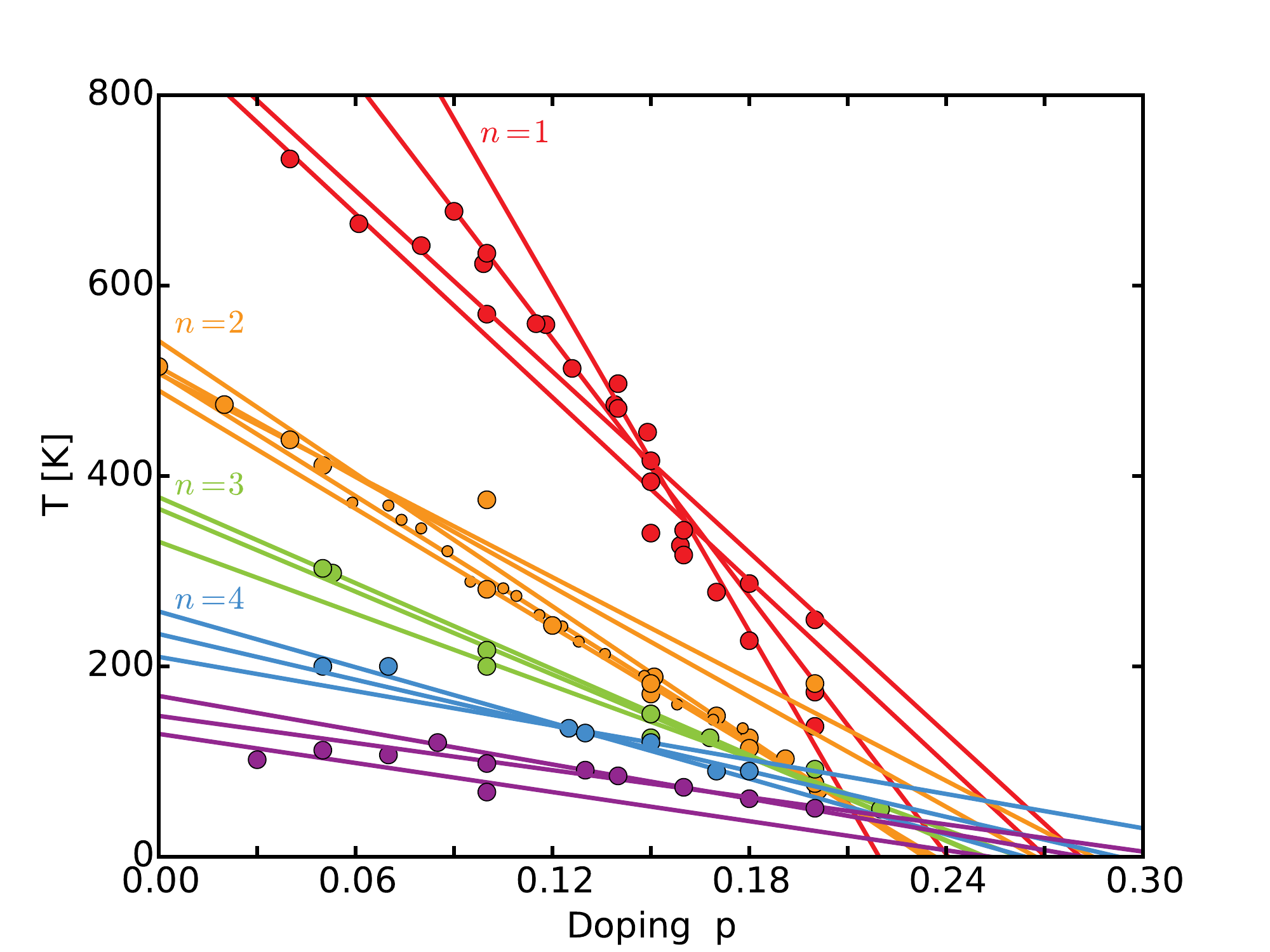}
\caption{ (Color online.)  Intersection of the pseudogap family in LSCO near $T=0,\,p=0.26$.   The straight lines show linear fits to individual pseudogap data sets, displayed as filled circles.   
}
\label{LSCOSlopes}
\end{figure}

The four characteristic temperatures just discussed follow a pattern which we have highlighted with thin black lines.   \footnote{We place the intersection at $p=0.26,\, T=0$ K  and align the second pseudogap line to the experimental data point in \cite{PhysRevB.46.14034} at $p=0, \; T=515$ K.    }  Both the experimental data and the black lines  are organized as harmonics of the uppermost line, with the $n$-th line having a  slope equal to $1/n$ times that of the highest $n=1$ line.  \footnote{ The average value of  $n \times dT/dp$, which is approximately constant because the pseudogap slope varies as $dT/dp \propto 1/n$, is  $4000 \,$ K, and the standard deviation of this value across the fifteen data sets in the first four pseudogap lines is $800\,$K.  
} Moreover all four lines radiate from  a point $p \approx 0.26 $ that  lies near the high-doping limit $p_{c2} = 0.27$ \cite{doiron2009correlation} of the superconducting dome. Figure \ref{LSCOSlopes} displays linear regressions of the eighteen pseudogap data sets which contain more than one data point.   The linear regressions of fourteen out of eighteen  pseudogap data sets intercept the $T=0$ axis in the interval $p=\left[0.23,0.29\right]$, of which six lie in $\left[0.25,0.27\right]$.   These regressions place the pseudogap's $T=0$ intercept on or near the high-doping edge of LSCO's superconducting dome, and clearly exclude the possibility suggested by some that the intercept lies well inside the dome, near  $p=0.19$. \cite{norman2005pseudogap}

Both the agreement of  intercepts and the pattern of $1/n$ slopes are remarkable given that we have used very broad selection criteria embracing an extremely diverse set of experimental techniques and sample preparation protocols, publications from 1990 to the present, and significant scatter within individual data sets.  While it is often suggested that the pseudogap temperature is sensitive to the experimental technique, we find instead that  all published measurements agree on the same simple pattern. 
 
These results force the conclusion that all four characteristic temperatures are members of the same family.  This family pervades most if not all of the cuprate phase diagram: in LSCO the $n=2$ pseudogap line, i.e. the transition to orthorhombic symmetry, persists to zero doping and $515$ K, and  the $n=1$ pseudogap line extends to above $700$ K. Since the $n=1,$  $n=3,$ and $n=4$ lines in this family are not structural transitions, the tetragonal to orthorhombic structural symmetry breaking on the $n=2$ line must be a subsidiary signal of  underlying electronic nematic order.

\section{YBCO data\label{YBCODataText}} 
\begin{figure}[]
\includegraphics[width=9.5cm,clip,angle=0]{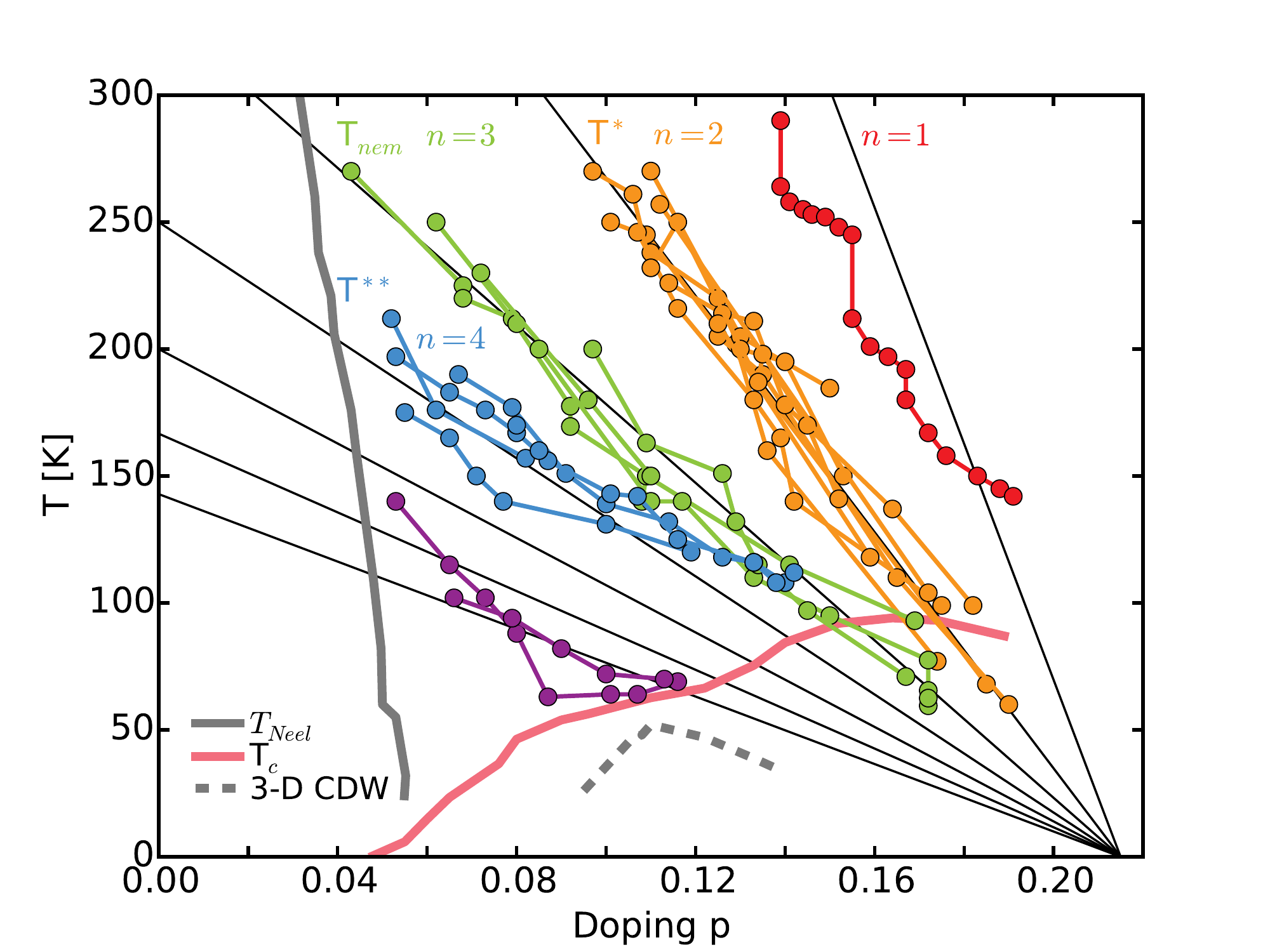}
\caption{ (Color online.) The  family of pseudogap lines in YBCO.  The black guides to the eye have slopes proportional to $1/n$ and meet at doping $p=0.215$.   Individual pseudogap data sets, each from a distinct experimental group measuring a distinct experimental probe and signature, are shown as filled circles connected by lines.   Near the first  $n=1$ pseudogap line a transport anomaly (red data set)  occurs.  $C_4$ and intra unit cell symmetries are broken on the second (orange) line $T^*$, while   on the third (green) line $T_{nem}$ transport becomes nematic and time reversal symmetry is broken.  The fourth pseudogap line $T^{**}$ (blue) is marked by transport signatures.  Purple lines show additional transport data sets. The pink line shows the superconducting $T_c$.  Underneath $T_c$ three-dimensional charge density wave order (dashed grey) occurs, and an antiferromagnetic phase (solid grey) is found at low dopings.
}
\label{YBCOData}
\end{figure}

Figure \ref{YBCOData} summarizes the experimental data on oxygen doped YBCO.  The experimental corpus contains a total of twenty-six distinct data sets from 1996 to 2018.  
Appendix \ref{YBCOAppendix} gives full details on each of these data sets, three more data sets which are in excellent agreement but which we have omitted because their data points show unusually large scatters \footnote{These sets are labeled in the appendix as  $n=2$ sets (e,k,l), and show scatters of $\pm 30$ K, which is substantially larger than the $\pm 15$ K shown by the other $n=2$ data sets.}, and another  data set which we have omitted on a case by case basis. \footnote{This is labeled in the appendix as Omitted Data Sets  (c).}

Although oxygen doped YBCO has a considerably higher maximum $T_c \approx 94$ K than LSCO, the data on its phase diagram is much curtailed. Hole dopings above $p=0.194$ cannot be obtained at ambient pressure, and the experimental literature on the pseudogap generally has not measured  dopings less than  $p \leq 0.05$, most likely because of proximity to the antiferromagnetic Neel phase, marked in grey in Figure \ref{YBCOData}.   Whether this practice assumes or confirms that the pseudogap lines do not coexist with Neel order seems unclear. We have not collated characteristic temperatures on the calcium doped variant of YBCO, in which hole dopings well in excess of $p=0.194$ can be achieved, because much less experimental data is available on characteristic temperatures in this variant compound.

In the remaining doping range  $ 0.06 \leq p \leq 0.194 $ we are able to clearly identify the $n=2,$ $n=3,$ and $n=4$ pseudogap lines, which follow rays intersecting near $p \approx 0.215$.    \footnote{We align the $n=2$ pseudogap line to $p=0,\,T=500$ K and $p=0.215, \, T=0$. }  The scatter of data around each characteristic temperature is about the same as in LSCO, of order $ 15$ K.  We find again  clear gaps between the characteristic temperatures, allowing each data set to be unambiguously assigned to one of the  characteristic temperatures.

 The  orange lines (ten data sets) show YBCO's pseudogap temperature $T^*$, marked by transport signatures and the onset of fluctuating intra unit cell order. Recent symmetry-oriented experiments show that this is a nematic $C_4$ symmetry breaking transition similar to the symmetry breaking seen on LSCO's $n=2$  line.  However these experiments go beyond the neutron scattering and X-ray diffraction experiments performed on LSCO, and find that  $C_2$, mirror, and inversion symmetries are also broken in addition to $C_4$ symmetry breaking. \cite{shekhter2013bounding,zhang2018discovery,zhao2017global,sato2017thermodynamic,leridon2009thermodynamic,PhysRevLett.96.197001, PhysRevLett.118.097003, PhysRevB.78.020506,mangin2015intra,daou2010broken,PhysRevB.97.064502,PhysRevLett.93.267001,PhysRevMaterials.2.024804,sidis2007search,PhysRevB.59.1497,alloul2010superconducting,wang2017revisiting,PhysRevLett.112.147001} 
On the  green $T_{nem}$ lines  (seven data sets), transport becomes nematic and time reversal symmetry is broken.   \cite{dai1999magnetic,kapitulnik2009polar,PhysRevB.92.224502,PhysRevB.53.9418,PhysRevLett.93.267001}  The nematic temperatures $T_{nem}$ have values equal to two thirds those of the pseudogap temperatures $T^*$, not one half.  This is the reason why we assign $n=2$ to $T^*$ rather than $n=1$, and $n=3$ to $T_{nem}$ rather than $n=2$.     At large dopings both the $n=2$ $T^*$ and $n=3$ $T_{nem}$ lines clearly cross the superconducting dome and coexist with the superconducting phase, excluding the possibility that the pseudogap transition merges with the superconducting phase transition rather than crossing it. \cite{norman2005pseudogap}  Next the  blue $T^{**}$  lines ($n=4$, six data sets) have been measured in resistance and Hall resistance experiments, and their values are one half those of the pseudogap temperatures $T^*$, leading us to assign $n=4$ to the  $T^{**}$ lines.\cite{PhysRevB.83.054506,PhysRevB.69.104521,PhysRevMaterials.2.024804}  At dopings $p < 0.085$ intra unit cell ordering occurs on this  $n=4$ pseudogap line rather than on the $n=2$   line. \cite{haug2010neutron,PhysRevB.83.104504,sonier2001anomalous}

 Turning to YBCO's $n=1$ pseudogap line,  experiments have almost never explored temperatures above $300$ K because of concerns about thermal history memory,   aging and equilibration, and preparation protocol.
  This leaves only a small corner  in the phase diagram  where the $n=1$ line might be seen, stretching from $p=0.15, \,T=300$ K to $p=0.194,\, T=95$ K, which limits our ability to ascertain the existence of a line here.   In this region there is  a large step change in the resistance and an onset of thermal history memory \cite{lavrov1992low, PhysRevLett.85.2376}.  Moreover Ando's resistance measurements  show a prominent white contour, plotted in red in Figure \ref{YBCOData},  which is roughly aligned with the expected $n=1$ line. \cite{PhysRevLett.93.267001}
  
\begin{figure}[]
\includegraphics[width=9.5cm,clip,angle=0]{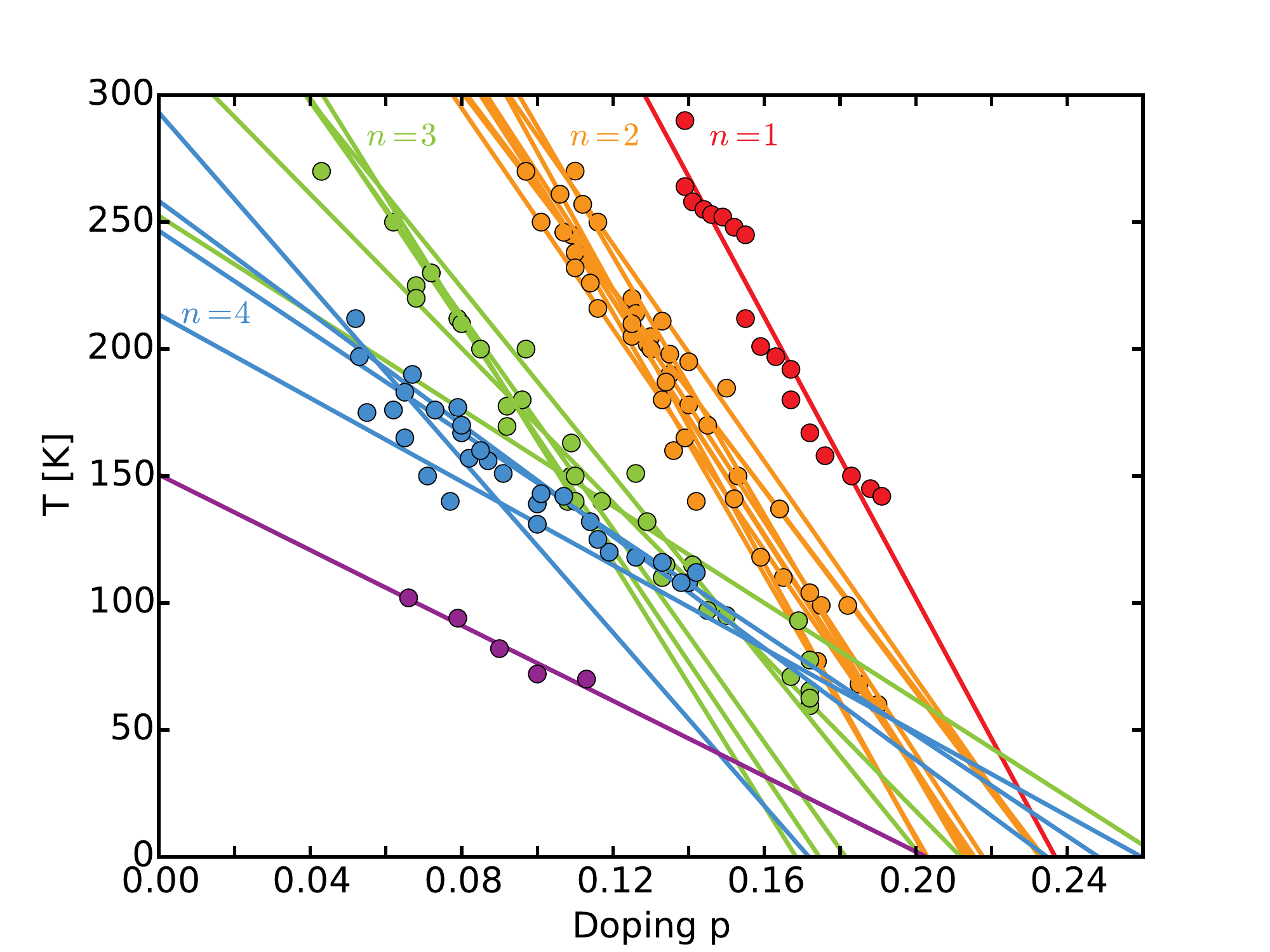}
\caption{ (Color online.)   Intersection of the pseudogap family in YBCO near $T=0,\,p=0.215$.   The straight lines show linear fits to individual pseudogap data sets, displayed as filled circles.  
}
\label{YBCOSlopes}
\end{figure}

  All four pseudogap lines intercept $T=0$  at  doping $p \approx 0.215$ with a scatter that is again notably small, comparable to the scatter around LSCO's $p\approx 0.26$. Figure \ref{YBCOSlopes} displays linear regressions of the twenty-two pseudogap data sets which contain more than one data point. \footnote{We have not included the regression of one of the low temperature (purple) data sets which clearly does not follow a linear trajectory, and instead seems to be composed of two line segments.} The linear regressions of fifteen out of twenty-two pseudogap data sets intercept the $T=0$ axis in the interval $p=\left[0.20,0.24\right]$, and six are clustered in $\left[0.21,0.22\right]$.    Only four out of twenty-two of the intercepts are found at dopings less than $p=0.20$, excluding  the possibility that YBCO's pseudogap intercept lies near $p=0.19$. \cite{norman2005pseudogap}


\section{Comparison} Comparing LSCO to YBCO, the pseudogap lines extrapolate to zero-doping temperatures which are the same within experimental error,  i.e. $515$ K for LSCO's $n=2$ line vs. $500$ K for YBCO. \footnote{It may be worth noting that at $p=0$ the $n=1$ line's temperature $T=1000$ K, if converted to area $A$ via $ \hbar^2 A^{-1} / 2 m_e  =   k_B T$, gives an area $A = 158\, a_0^2$.  This is three times the area of the copper oxide unit cell. } However the pseudogap family's meeting point on the $T=0$ axis lies at considerably different dopings: in  YBCO it lies at $p \approx 0.215$,  while in LSCO it is found at $p \approx 0.26$.  This should lay to rest  debates about the $T=0$ intercept of the pseudogap line, and about the location of the putative quantum critical point: if the question is framed  in terms of hole doping $p$, the answer  clearly depends on the material.   Yet the question can be answered in a different way that is not material dependent: in both LSCO and YBCO  the pseudogap's $T=0$ intercept lies very near to the maximal doping $p_{c2}$ at which superconductivity can be achieved, which has two different values, $p_{c2} = 0.27$ \cite{doiron2009correlation} in LSCO vs. $p_{c2} = 0.194$ \cite{PhysRevB.73.180505} in oxygen-doped YBCO at ambient pressure.   In this sense the pseudogap lines in both LSCO and YBCO reach $T=0$ at the real material-dependent end of the superconducting phase.  We conclude that in both materials the entire superconducting dome  lives under the shelter of the pseudogap family, and that its overdoped edge is tied to the pseudogap.

The various observables marking  the pseudogap family collapse at critical dopings which depend on the line,  and the doping scheme \cite{PhysRevB.95.224517}, and the observable. On LSCO's $n=1$ line the linear resistance signature  collapses between $p=0.17$ and $p=0.18$ \cite{PhysRevB.49.16000, cooper2009anomalous, PhysRevB.97.064502}, while the peak in the magnetic susceptibility continues past $p=0.20$ \cite{PhysRevB.49.16000}.   On the $n=2$ line the tetragonal phase associated with this line collapses near $p=0.208$ \cite{PhysRevLett.68.3777}, and on the $n=3$ line the ARPES pseudogap persists until at least $p=0.22$.   \cite{PhysRevB.75.140503}.      
It seems likely that the collapse of each line is caused by a small competing temperature scale, since in both LSCO and YBCO the $n=2,\,n=3,$ and $n=4$ lines either collapse or disappear from the experimental record between $50$ and $110$ K. The coupling between copper oxygen planes is a likely candidate for supplying the competing scale, since it supports long range 3-D charge density waves and superconductivity in roughly the same temperature range.

The experimental signatures associated with each of the pseudogap lines in YBCO could differ substantially from the signatures seen on corresponding lines in LSCO.   Except for  the $n=2$ line, comparisons of symmetry breaking in LSCO and YBCO  are impossible because LSCO pseudogap experiments have generally not probed symmetry breaking.  Moreover, the hallmark of  LSCO's $n=3$ line is that the ARPES density of states begins to manifest a pseudogap, while in YBCO this change  is generally believed to  occur instead on YBCO's "pseudogap line" i.e. its $n=2$ line.   Since there is no ARPES data on YBCO's pseudogap temperature, it remains possible that  in YBCO the  pseudogap could manifest in the density of states on the $n=3$ line rather than the $n=2$ line, the same as in LSCO.    This question is confused further by data on $p\approx0.14$ Bi$_2$Sr$_2$CaCu$_2$O$_{8+\delta}$ (Bi2212)  where  the ARPES pseudogap seems to open in two discrete steps near $T=250 $ K and $T = 150$ K, suggesting that in Bi2212 the density of states may change substantially at two distinct pseudogap lines.  \cite{vishik2018photoemission}

\section{Analysis}  These results are very fertile ground.   First, the fact that the pseudogap family spans all dopings up to $p_{c2}$ and temperatures up to $700$ K argues strongly that the entire phase diagram up to $p_{c2}$  hosts a single mother phase or order.  The various phenomena observed along the pseudogap lines, for instance the structural phase transition to orthorhombic order,  are then subsidiary or parasitic phenomena that respond to underlying changes in the mother phase.  The details of which line a particular observable (such as orthorhombic symmetry) is associated with may depend on the material, and even when a particular observable  in a particular material  collapses at a particular doping the underlying line in the mother phase may continue robustly to higher dopings.   As a case in point, in YBCO intra unit cell order switches from the $n=4$ line to the $n=2$ line near $p=0.085$ \cite{haug2010neutron,sonier2001anomalous,PhysRevLett.96.197001, PhysRevLett.118.097003, PhysRevB.78.020506,PhysRevB.83.104504,mangin2015intra,sidis2007search}, while transport signals show that the $n=4$ line continues until at least $p=0.142$.  \cite{PhysRevMaterials.2.024804,PhysRevB.83.054506,PhysRevB.69.104521}   
Superconductivity may also be a subsidiary or parasitic phenomenon which occurs when (a) the mother phase assists hole transport and (b) the interlayer coupling is strong enough to support long range 3-D order.

Secondly, the width of the pseudogap lines is unquestionably sharp compared to the pseudogap temperatures themselves.  Figures \ref{LSCOData} and \ref{YBCOData} show that the scatter around each pseudogap line is typically of order $\pm 15$ K. 
\footnote{Judging from authors' estimates,  comparison of data sets to their linear regressions, and comparison between data sets, most data sets show scatters or errors of $\pm10$ to $\pm20$ Kelvin.  A notable exception is \cite{shekhter2013bounding}, which reports that YBCO's $n=2$ pseudogap line has a width of $3$ K. There are also several data sets with exceptionally large scatters of $50$ K or more, including three out of four data sets on LSCO's $n=1$ line and also the thermoelectric power data on LSCO's $n=2$ line. \cite{yoshizaki1990magnetic,PhysRevB.49.16000,kim2004two}  We have not included in our figures three data sets on YBCO's $n=2$ pseudogap line because their scatters are larger than $15$ K \cite{leridon2009thermodynamic,PhysRevB.59.1497} , including   neutron scattering data revealing intra unit cell order. \cite{PhysRevLett.96.197001, PhysRevLett.118.097003, PhysRevB.78.020506,PhysRevB.83.104504,mangin2015intra,sidis2007search}} 
 This sharpness controverts the hypothesis that the pseudogap line is a crossover, and could be understood as evidence of standard phase transitions.  However the fact that there is a family of  repeated pseudogap lines indicates that we are instead seeing a quantum coherent effect in the mother phase, in the same category as Landau levels or atomic orbitals. If so, then quantum coherence must persist to temperatures as high as $700$ K in the cuprates.  This conclusion is reinforced by parallel evidence  from  the strange metallic (linear in magnetic field and linear in temperature) resistance seen in the cuprates, which is also a manifestation of quantum coherence at temperatures far above $T_c$. \cite{hayes2016scaling,giraldo2018scale,sacksteder2018fermion}

Thirdly,  the pseudogap lines are in truth linear.  Clarity about this linearity was obtained by restricting our data to single materials with a single doping scheme.  There are some mild deviations from linearity  - YBCO's $n=4$ pseudogap line flattens at around $140$ K, and LSCO's $n=1$ and $n=2$ lines steepen  at high dopings - but these are mild distortions, and are the exception rather than the rule.  
 The pseudogap lines relate temperature $T$ to the 2-D sheet density of holes $\rho_{holes} = p / \mathcal{A}$  within the copper oxide plane.  Here $\mathcal{A}$ is the area of the copper oxide unit cell. Since in atomic units sheet density has the same units as both temperature and energy, we conclude that the pseudogap temperatures are direct measures of a sheet density.  This conclusion is supported by the fact that in atomic units the constant of proportionality between   the $n=1$ pseudogap line and $\rho_{holes}$ is of order one: $-1.27$ for LSCO and $-1.55$ for YBCO. 
The natural endpoint of this reasoning is that in the cuprates there is a two dimensional sheet density $\Pi_{psg}$ which controls both the pseudogap family and the family's mother phase, that $\Pi_{psg}  \propto p_{c2}  - p$ is high at low dopings and decreases linearly to zero at $p_{c2}$, and that the pseudogap temperatures are direct measures of this density.

Two other linear relations between sheet density and temperature have already been seen  in the cuprates.  Uemura's proportionality relation between the superfluid  density and $T_c$ holds in many underdoped materials \cite{PhysRevLett.62.2317}, and a formally identical relation has been verified also in overdoped LSCO \cite{bozovic2016dependence}.  Secondly, several LSCO and Bi2212 experiments have reported temperature scales which rise linearly with hole doping $p$ and extrapolate to zero at the underdoped end of the superconducting dome, so that these temperatures seem to measure the sheet density of mobile holes - see Figure \ref{CoherenceTemperatures} in Appendix \ref{CoherenceTemperature} for a summary of these results.  \cite{kim2004two,ohsugi1991cu,PhysRevB.79.140502,chatterjee2011electronic} It is also significant that 3-D long range charge density wave (CDW) order and the $1/8$-th anomaly occur near  half-filling for $\Pi_{psg}$, suggesting that these may be favored when the pseudogap density is about one half of its maximum value.  

It is technically possible that the pseudogap density $\Pi_{psg}$ and pseudogap lines could reflect strictly ultraviolet i.e. short range physics near the atomic scale, and that the cuprate phase diagram is not controlled directly by $\Pi_{psg}$ but instead through renormalization group flow from short range to long range collective behavior.  This has the weaknesses that atomic scale physics would be expected to cause resonances in the phase diagram at values of $\Pi_{psg}$ tied to the crystal structure, but these are not seen, and that at dopings near $p_{c2}$  the area scale indicated by $\Pi_{psg}^{-1}$ is far in excess of the atomic scale.   In our view it is more likely that  $\Pi_{psg}$ is a fundamental determining property of long-range order in the pseudogap mother phase.  We leave for future work the question of what quantity is counted by $\Pi_{psg}$, although some of the obvious possibilities are vortices, skyrmions, dislocations, entanglement density, or topological quantities.

       \begin{figure}[]
\includegraphics[width=9.5cm,clip,angle=0]{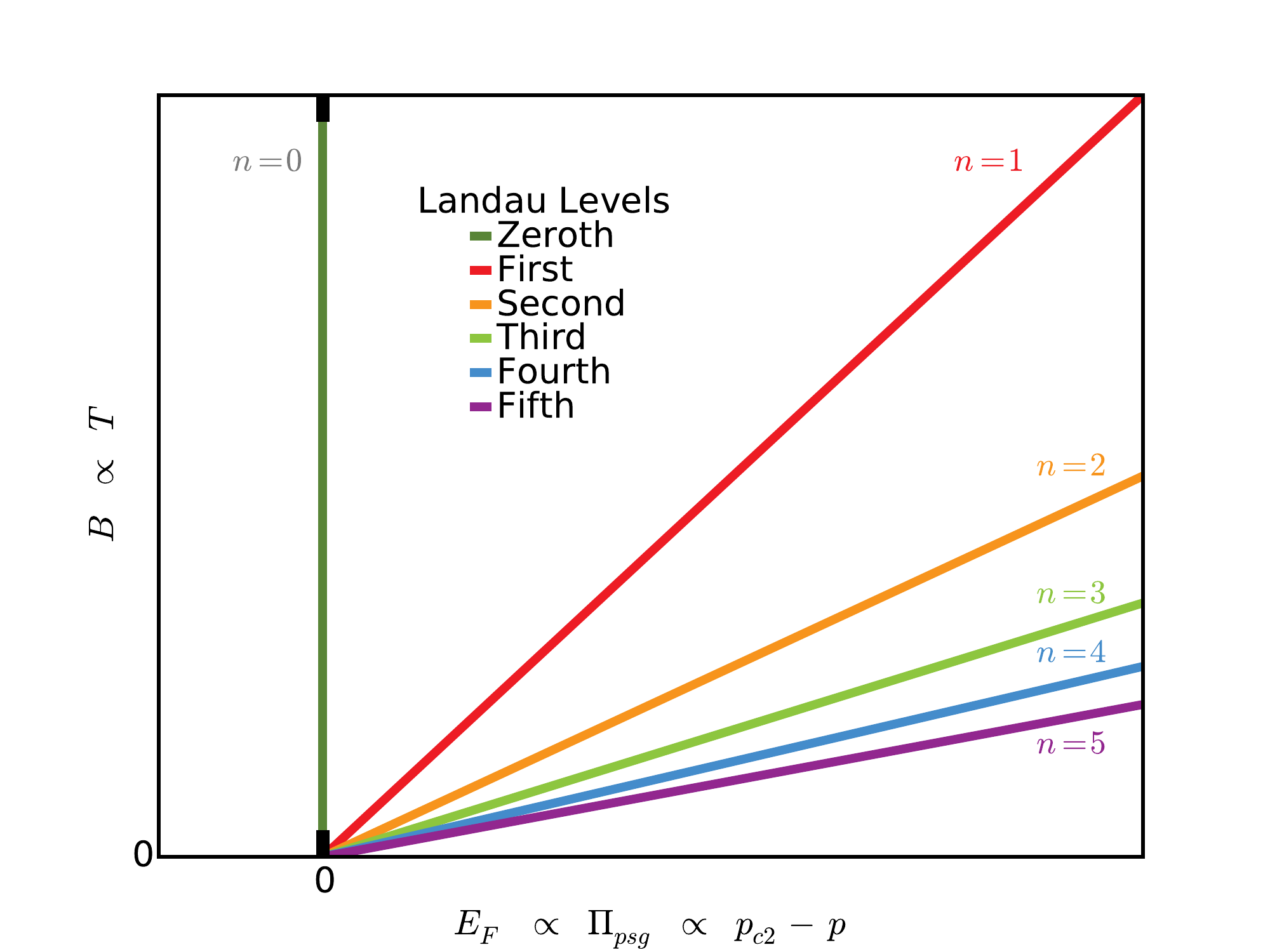}
\caption{ (Color online.) Effective model producing the same pattern seen  in the pseudogap family.   The model describes free two-dimensional $p^2/2m_e$ fermions with a Zeeman splitting $\mu_0 B \sigma_z$ moving in a magnetic field $B$.   We plot   the resulting Landau levels with their energies  $E_n$ along the $x$ axis and with magnetic field $B$ along the $y$ axis.  The Landau level slopes follow the same $1/n$ pattern seen in the pseudogap family if field $B$ is mapped to temperature and Fermi energy $E_F$ is mapped to pseudogap density $\Pi_{psg}$.
}
\label{IQHEdiagram}
\end{figure}

Fourthly, we turn to the $1/n$ quantization rule which controls the slopes.  In this connection we are inspired by recent experimental and theoretical  work which shows a direct linear relation between magnetic field $B$ and temperature $T$ in the strange metal phase of LSCO and other bad metals.  \cite{hayes2016scaling,giraldo2018scale,sacksteder2018fermion,kumar2018high}   The pattern traced by the pseudogap family  can be reproduced by a model of free two dimensional $p^2/2m$ fermions where  temperature is mapped to an effective magnetic field $B$ and  the pseudogap sheet density $\Pi_{psg} \propto p_{c2} - p$ is mapped to a Fermi level $E_F$.   As shown in Figure \ref{IQHEdiagram}, if  the effective model includes either a Berry phase \cite{PhysRevLett.82.2147} or a Zeeman term $\mu_0 B \sigma_z$, then its Landau levels have the same $1/n$ slopes seen in the pseudogap family and intersect at a common point $B=0, \;E_F=0$.   This model leads us to the conjecture that the pseudogap family is in some way related to the Integer Quantum Hall Effect.  Our conjecture is not a suggestion of well-defined quasiparticles or, going further, a Fermi liquid or  band structure. Topology can control  conduction even when  these concepts are not relevant, as has been seen in studies of strongly disordered topological insulators. \cite{PhysRevB.85.195140}


Figure \ref{IQHEdiagram} shows that  the effective model possesses  an additional $n=0$ Landau level at $E_F = 0$, which maps to a vertical line in the cuprate phase diagram at the upper critical doping $p_{c2}$.  This prediction is confirmed in LSCO by three experimental signatures which are extinguished on this line: the superfluid density \cite{bozovic2016dependence}, the electronic nematicity which is found at all lower dopings \cite{wu2017spontaneous}, and the coefficient of the linear-in-temperature contribution to the resisitivity \cite{doiron2009correlation,hussey2011dichotomy}.  


In summary, we have shown that the experimental corpus on cuprate characteristic temperatures shows that signals such as peaks, kinks, or extinction of diffraction peaks are found on four pseudogap lines which span the cuprate phase diagram.  In the large gaps between the four lines no such signal is found and temperature dependence is smooth.  This observation is also a prediction about future measurements of cuprate characteristic temperatures.  We have argued that the entire cuprate phase diagram from zero doping out to $p_{c2}$ hosts a single mother phase which is controlled by a two dimensional sheet density, and that the observed family of pseudogap lines are subsidiary phenomena caused by changes in the mother phase.  We also suggest that the superconducting phase and charge density wave order are supported by the mother phase and occur when it is augmented with an interplane coupling and, in the case of superconductivity, with hole carriers.




\clearpage

\appendix
\section{Temperature scales that rise proportionally to hole doping\label{CoherenceTemperature}}

\begin{figure}[]
\includegraphics[width=9.5cm,clip,angle=0]{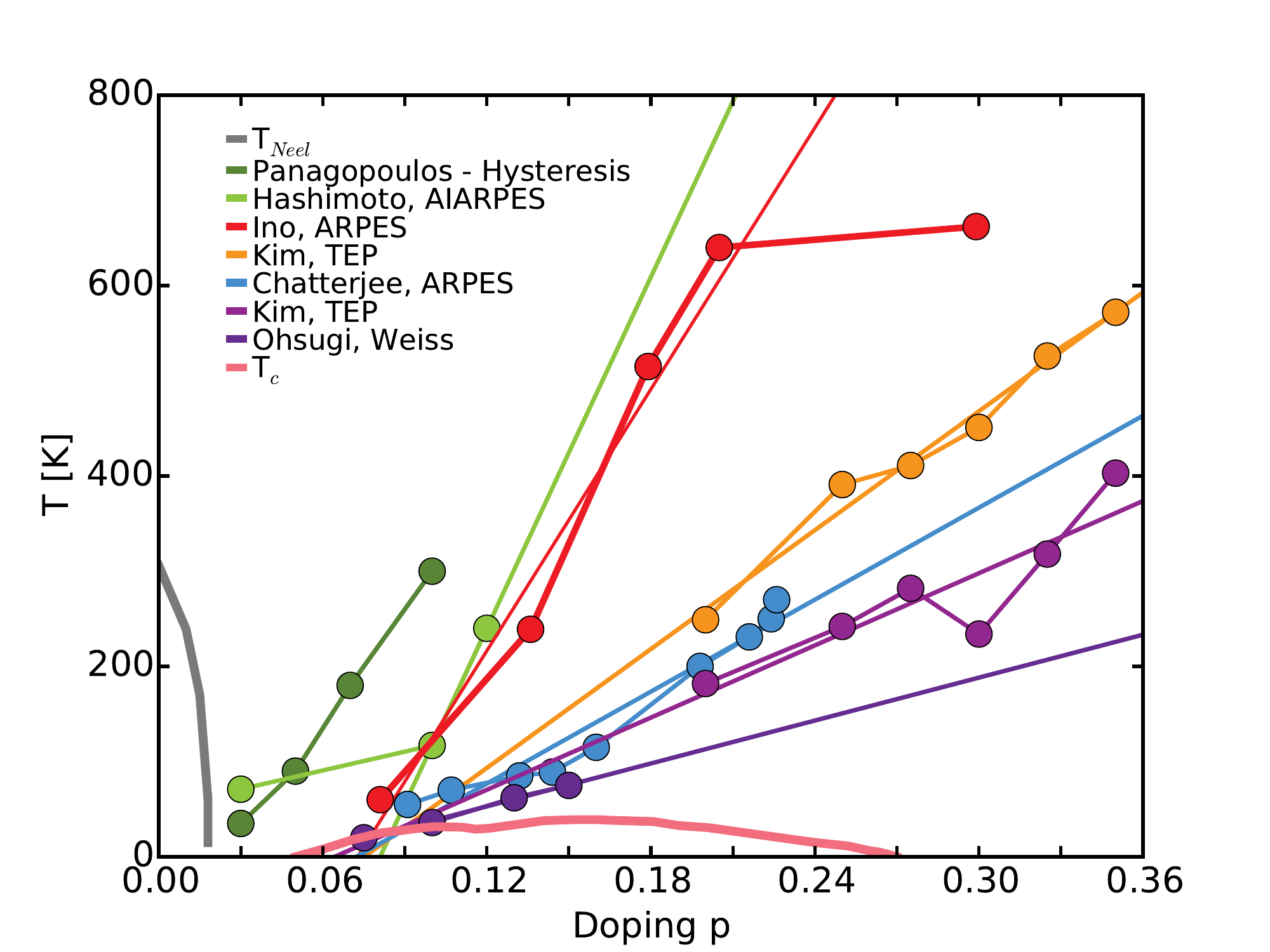}
\caption{ (Color online.) Characteristic temperatures  that seem to be direct manifestations of the density of mobile holes.  Light green, red, and blue data sets were measured using ARPES, yellow and light purple sets with the thermoelectric power, and the dark purple data set with the magnetic susceptibility. 
The dark green set measuring the onset of magnetic hysteresis intercepts the $p$ axis at lower doping, perhaps because it is sensitive to pinned holes. The superconducting dome is shown in pink, and the antiferromagnetic phase in grey. 
}
\label{CoherenceTemperatures}
\end{figure}

With the exception of Refs. \cite{PhysRevLett.81.2124,PhysRevLett.96.047002}, the  data sets graphed in Figure \ref{CoherenceTemperatures} are temperatures that rise linearly with doping and extrapolate to $T=0$ near the underdoped limit of the superconducting dome, i.e. in the interval $p = \left[0.049,0.081\right]$. This suggests that these temperatures are direct manifestations of the sheet density of mobile holes. Ref. \cite{PhysRevLett.81.2124} reports an energy scale with a similar behavior.  

Ref. \cite{PhysRevLett.96.047002} also reports a temperature which rises linearly up to $T = 300$ K.  This last data set extrapolates to  $T=0$ at a lower doping $p=0.023$, presumably because of a sensitivity to pinned holes.

All data sets used in Figure \ref{CoherenceTemperatures}, with discussion of their particulars and origin, and with a script that produces this figure, are available in the supporting material as a python script.

\begin{enumerate} 
\item Hashimoto, 2009. \cite{PhysRevB.79.140502} Angularly integrated ARPES. LSCO. The authors take the first derivative of the spectrum with respect to energy, and identify a peak in the first derivative.  The temperature reported here, which they call a coherence temperature, is a break in the temperature dependence of the peak position. We omit a data point at $p=0.15$ which fits well with the linear regression because they did not actually reach the reported temperature $T = 364$ K.    Roughly consistent with Ref. \cite{PhysRevLett.81.2124}. The slope is $6200$ K without the $p=0.15$ data point, or $4900$ K with it. 
\item Ino, 2009. \cite{PhysRevLett.81.2124} Angularly integrated ARPES. LSCO. The quantity reported here is a measure of  the width of the Fermi surface. Unlike all other data discussed in this article, this is an energy scale converted to temperature, not an experimental temperature.  The experiment was performed at $T =  18$ K.  The quantity reported here includes a factor of $1/\pi$ which might be able to be renormalized at will. We include this data set because the four data points between $p=0.074$ and $p=0.203$ rise linearly with doping and extrapolate to the underdoped edge of the superconducting dome.  We omit the $p=0.30$ point from the linear fit, and we omit the $p=0$ point altogether because Figure 2 in Ref. \cite{PhysRevLett.81.2124} shows $p=0$ data that seems to leave little ground for extracting a width. The slope is $4600$ K. 
\item Kim, 2004.  \cite{kim2004two} Thermoelectric power.  LSCO. The temperature reported here marks a  break from the linear signal seen at high temperatures.  Here we plot only the dopings at $p=0.20$ and higher.   The slope is $2100$ K, about half of the slope in Ref. \cite{PhysRevLett.81.2124}. 
\item Chatterjee, 2011. \cite{chatterjee2011electronic} ARPES. Bi2212. Below this temperature the spectrum contains a sharp Gaussian peak, and above this temperature the peak is absent.  Unlike all other data discussed in this article, this data is obtained from Bi2212. 
\item Kim, 2004. \cite{kim2004two}  Thermoelectric power. LSCO. The temperature reported here marks a break from the  linear signal seen at low temperatures.  Here we plot only the dopings at $p=0.20$ and higher.  
\item Ohsugi, 1991.  \cite{ohsugi1991cu}   Nuclear quadrupole resonance. LSCO.  The temperature reported here is a Weiss temperature obtained by fitting the nuclear spin relaxation rate to a Curie-Weiss law. 
\item  Panagopoulos, 2006.   \cite{PhysRevLett.96.047002}  See also Ref. \cite{PhysRevB.69.144508} by the same group. LSCO. This temperature marks the  onset of hysteresis in the temperature dependence of the low field magnetization, which is probably a sign of pinned vortices and of pairing. The low-doping data from $p=0.03$ to $p  = 0.10$ nicely follows a straight line originating at $T=0, \, p=0.023$ and extending up to room temperature.  The small $p=0.023$ intercept may be caused by the observable's sensitivity to both pinned and mobile holes.  The slope is about $3900$ K, roughly comparable to the slopes of the LSCO ARPES data sets. \cite{PhysRevB.79.140502,PhysRevLett.81.2124}
\end{enumerate}

\section{LSCO Data Sets\label{LSCOAppendix}}
 The temperatures gathered here were all realized experimentally rather than by extrapolation from lower temperatures, and clearly identifiable signals  occurred at the reported temperatures, including peaks, kinks, extinction of diffraction peaks, etc.   In the interest of clarity we do not rely on universality arguments, and therefore restrict ourselves to lanthanum cuprate with strontium doping(LSCO), and we  keep separate the results of distinct experimental groups and of distinct experimental probes and signatures. 
We do not report any temperatures that are not already reported by the articles we have cited.  In particular, we have stayed out of the business of re-analyzing or fitting data sets from other articles.  The one exception to this rule is our use of the color maps in \cite{PhysRevLett.93.267001} - from that article we extracted data from certain contours and features that are prominent in the color maps.

 Our survey of pseudogap temperature measurements does not extend to the extensive  literature on anomalies and phase transitions measured using mechanically-oriented observables such as internal friction, sound velocity, lattice constants, thermal expansivity, and the like. \cite{dominec1993ultrasonic} 

All data sets used in the LSCO figures or enumerated here, with discussion of their particulars and origin, and with a script that produces the figures, are available in the supporting material as a python script.

\begin{enumerate}
\item $n=1$ line: In LSCO the $n=1$ line marks a resonance in the magnetic susceptibility and qualitative changes  in transport and in the thermoelectric power.
\begin{enumerate}
\item Yoshizaki, 1990. \cite{yoshizaki1990magnetic} Peak in the magnetic susceptibility.  
\item Nakano, 1994. \cite{PhysRevB.49.16000}  Peak in the magnetic susceptibility.  The peak continues past $p=0.20$, and disappears near $p=0.22$.  We omit the last three data points because no peak exists in those susceptibility curves until a Curie term has been subtracted.  We also omit the first four data points, at lowest doping, because the experimental temperature did not go high enough to see the actual peak.  The scaling analysis used to obtain those four points is however very convincing, especially because the peak was seen at lower dopings by \cite{yoshizaki1990magnetic}.  
\item Kim, 2004.  \cite{kim2004two} Thermoelectric power.   The temperature reported here marks a  break from the linear signal seen at high temperatures.  Here we plot only the dopings up to $p=0.20$.  
\item Nakano, 1994. \cite{PhysRevB.49.16000}   Resistivity.   The temperature reported here marks a break  from the linear signal seen at high temperatures.  The break disappears and the signal seems to be perfectly linear (from visual inspection) at $p=0.16$ and $p=0.18$, which is consistent with Ref. \cite{cooper2009anomalous}'s data at $p=0.17,\,0.18$. This is also consistent with Ref. \cite{PhysRevB.97.064502}'s statement that the resistivity signal of the pseudogap collapses between $p=0.17$ and $p=0.18$. 

 There is apparent disagreement between Ref. \cite{PhysRevB.49.16000} and  Ref. \cite{laliberte2016origin}, which was cited by Ref. \cite{PhysRevB.97.064502} by the same authors.  Ref. \cite{PhysRevB.49.16000} reports that at $p=0.14$ and $p=0.16$  the pseudogap temperature, i.e. the onset of linear in temperature resistance, occurs at $471$ K and $317$ K respectively.  In contrast  Ref. \cite{laliberte2016origin} gives resistivity data at $p=0.136,\, 0.143,\, 0.157,\, 0.163$ that superficially indicates much smaller temperatures.  The $p=0.136$ resistivity seems to be linear above $150$ K or so, and the $p=0.143$ resistivity seems to be linear above $110$ K.  However  Ref. \cite{laliberte2016origin}'s data goes up to only $200$ K, so they are unable to detect any linearity above $471$ K.  In contrast, Ref. \cite{PhysRevB.49.16000}'s data at $p=0.14$ extends up to $900$ K. Moreover, Ref. \cite{PhysRevB.49.16000}'s  data  extends down to the superconducting temperature, shows that the slope below $471$ K is not very different from the slope  above that temperature, and changes only gradually.  Therefore Ref. \cite{laliberte2016origin} may just not have enough data to detect nonlinearity near the high end of their temperature range.  A second alternative is that there could be two linear regimes, one above $471$ K, and another above $110$ - $140$ K, corresponding to two pseudogap temperatures.  
\end{enumerate} 
\item $n=2$ line: In LSCO the $n=2$ line is a symmetry breaking transition from $C_4$ down to $C_2$ nematic order. It has been commonly regarded as a structural transition from tetragonal to orthorhombic symmetry.  It is accompanied by features in transport and in the thermoelectric power.
\begin{enumerate}
\item Kim, 2004. \cite{kim2004two}  Thermoelectric power. The temperature reported here marks a break from the linear signal seen at low temperatures. 
\item Takagi, 1992. \cite{PhysRevLett.68.3777} X-ray diffraction, looking for a peak splitting caused by orthorhombicity.   This data shows that the signal i.e. the orthorhombic phase collapses near $p=0.208$. Our linear fit omits  two points dropping almost vertically near p=0.208.  
\item Ando, 2004. Resistivity. \cite{PhysRevLett.93.267001}  We reproduce a red linear feature that is very prominent in Ando's plot.  The feature disappears in the range between $p=0.178$ and $p=0.19$. 
\item  Yamada, 1998.  \cite{PhysRevB.57.6165}  Neutron scattering sensitive to orthorhombic symmetry. 
\item Keimer, 1992. \cite{PhysRevB.46.14034} Neutron scattering sensitive to orthorhombic symmetry.  More specifically, extinction of the $(021)$ superlattice reflection peak.  
\end{enumerate}
\item $n=3$ line: In LSCO the  $n=3$ line, also called the pseudogap temperature, is marked by the birth of a pseudogap controlling the density of states.   It is accompanied by features in transport and in the thermoelectric power.
\begin{enumerate}
\item Yoshida, 2012.  \cite{PhysRevLett.103.037004}  ARPES.  The data here is the temperature where the pseudogap (measured with ARPES) disappears.
\item Ando, 2004. \cite{PhysRevLett.93.267001}   Resistivity.  We reproduce a line segment imposed on the data by the authors which follows qualitative trends of the colors in their graph.  We omit the first of the two line segments which they imposed, a constant-temperature $T=298$ K  line at low doping $p \leq 0.053$.  This $T=298$ K line is in fact supported by colors in the graph.  At the overdoped end, the authors' line continues to $p=0.168, \, T = 125$ K, but the underlying resisitivity data suggests that the line should stop a little earlier, between $p=0.15$ and $p=0.168$.  
\item Kim, 2004.  \cite{kim2004two} Peak in the thermoelectric power.     Here we plot only the dopings from $p=0.05$ up to $p=0.20$. At higher dopings the signal increases very slowly with doping. 
\item Hashimoto, 2007. \cite{PhysRevB.75.140503}  Angularly integrated ARPES.  At temperatures below this temperature the density of states increases relatively rapidly with temperature, and above this temperature the DOS increases less quickly with temperature - i.e. the slope changes from one value to another in a discontinuous way. We omit a zero temperature data point at p=0.30 because it only bounds the doping value  where this line collapses to the range between $p=0.22$ and $p=0.30$.  We also omit the $p=0.03, \; T=300$ K data point because here no break is observed in the data, i.e. there is no data above $300$ K.  This pseudogap temperature is based on the authors noticing that at other dopings the slope below $T^*$ decreases inversely with $T^*$, as seen in the inset of their Figure 2a.  This  $p=0.03$ data point agrees very well with all other data points on this line, and extends the line down to $p=0.03$. 
\item Matt, 2015. \cite{PhysRevB.92.134524} ARPES.  Based on measurements of the normal-state antinodal spectral gap - $T^*$ is the temperature where this gap goes to zero. This data point is omitted because it concerns Nd-LSCO, but it does match the $n=3$ line well.  They also show a strong pseudogap at $p=0.12$ which results in a bound $T^* > 75$ K, though probably $T^*$ is much larger than $75$ K. 
\end{enumerate}
\item $n=4$ line: In LSCO the $n=4$ line is marked by features in transport and in nuclear magnetic resonance.

We plot only the first three data sets listed here.   The same pseudogap line revealed by these three data sets is very well supported by resistivity and Nernst data from Nd-LSCO and  Eu-LSCO, as documented in the remaining data sets. All resistivity and Nernst data sets are shown together in   Ref. \cite{PhysRevB.97.064502} and show impressive agreement.
\begin{enumerate}
\item Cyr-Chinoire, 2018 re-analyzes Fujii, 2010. \cite{PhysRevB.97.064502,FUJII2010S21}  Nernst effect.  The temperature reported here marks a  break from the linear signal seen at high temperatures. 
\item Cyr-Chinoire, 2018 re-analyzes Ong, 2010/2011.  \cite{PhysRevB.97.064502,xu2000vortex,PhysRevB.64.224519} Nernst effect.  The temperature reported here marks a  break from the linear signal seen at high temperatures.  This data set  is a re-analysis of Nernst effect data  from two papers which have Wang, Xu, Ong, and Uchida as co-authors.    
\item Itoh, 2004.  \cite{PhysRevB.69.184503} NMR.  Peak in the nuclear spin-lattice relaxation rate. 
\item Cyr-Choiniere, 2018. \cite{PhysRevB.97.064502,FUJII2010S21,cyr2009enhancement} Nernst effect.  The temperature reported here marks a  break from the linear signal seen at higher temperatures. This data set of three data points from $p=0.15$ to $p=0.21$ is omitted because it concerns Nd-LSCO, but it matches the LSCO $n=4$ line well.  It includes one data point from Ref. \cite{FUJII2010S21}.  It also includes two data points from Cyr-Choiniere's Ref. \cite{cyr2009enhancement}.  A third $p=0.24$ data point from Cyr-Choiniere is not included in the data set (a double omission) because it  lies at $T=0$ and therefore it only bounds the doping value rather than fixing it - however this data point does show that in Nd-LSCO this pseudogap line collapses somewhere between $p=0.21$ and $p=0.24$.  
\item  Collignon, 2017. \cite{PhysRevB.95.224517} Resistivity.  The temperature reported here marks a  break from the linear signal seen at higher temperatures.  This data set from $p=0.20$ to $p=0.24$ is omitted because it concerns Nd-LSCO, but it matches the $n=4$ line well.  When fitting to a straight line, we (doubly) omit a $T=0$ data point at $p=0.24$.  The pseudogap temperature at $p=0.23$ is $40$ K, so the pseudogap temperature collapses to zero between $p=0.23$ and $p=0.24$. 
\item  Collignon, 2017, analyzes Ichikawa, 2000. \cite{PhysRevB.95.224517,PhysRevLett.85.1738} Resistivity.  The temperature reported here marks a  break from the linear signal seen at higher temperatures.  This data set from $p=0.12$ to $p=0.15$ is omitted because it concerns Nd-LSCO, but it matches the $n=4$ line well.  This data set is two data points obtained by re-analyzing data from Ichikawa.  
\item Cyr-Chinoire, 2018. \cite{PhysRevB.97.064502,cyr2009enhancement} Nernst effect. The temperature reported here marks a  break from the linear signal seen at higher temperatures.  This data set from $p=0.08$ to $p=0.21$ is omitted because it concerns Eu-LSCO, but it matches the LSCO $n=4$ line well. One data point within this data set is obtained from re-analysis of work from separate authors, so we (doubly) omit it.  Three data points are from Ref. \cite{PhysRevB.97.064502} and one data point is from   Ref. \cite{cyr2009enhancement} by the same authors.  
\end{enumerate}
\item Other Data Sets:
\begin{enumerate} 
\item Baledent, 2010.  \cite{PhysRevLett.105.027004} Neutron scattering detecting intra unit cell two-dimensional short range order.  This data point lies 4K above the $n=6$ line, well within experimental error bars. 
\item Itoh, 2004. \cite{PhysRevB.69.184503} NMR.  A minimum in the nuclear spin-lattice relaxation rate. This data set looks like it could belong to the 5th or $n=6$ line. 
\item Xu, 2000. \cite{xu2000charge} Tangent of the Hall angle.  The temperature reported here marks a  break from a quadratic form which gives a good fit at higher temperatures. We omit a data point which lies at $T=0, \; p=0.17$ because it only bounds the doping value where this temperature goes to zero to the interval between $p=0.10$ and $p=0.17$.  
\item Matsuzaki, 2004. \cite{matsuzaki2004electronic}  Peak in the heat capacity divided by temperature.  This data set lies near the $n=5$ and $n=6$ lines.  
\end{enumerate}
\item Superconducting $T_c$ and Neel temperature:
\begin{enumerate} 
\item Momono, 1994. \cite{momono1994low}   $T_c$. 
\item  Ando, 2004. \cite{PhysRevLett.93.267001} $T_c$. 
\item Doiron-Leyraud, 2009. \cite{doiron2009correlation} $T_c = 0$ at $p= p_{c2}=0.27$.
\item Keimer, 1992. \cite{PhysRevB.45.7430}  Neel temperature measured with neutron scattering. 
\item Matsuda, 2002. \cite{PhysRevB.65.134515}  Neel temperature measured with neutron scattering. 
\item Niedermayer, 1998. \cite{PhysRevLett.80.3843}  Neel temperature measured with muon spin rotation.  
\end{enumerate}
\item Omitted Data Sets:
\begin{enumerate}
\item Panagopoulos, 2004, 2005, 2006. \cite{PhysRevB.69.144508,PhysRevB.72.024528,PhysRevLett.96.047002} Onset of hysteresis in the temperature dependence in the low field magnetization. We omit these data sets because they are clearly not related to the pseudogap family. They are look like a high-temperature replica of the superconducting dome, reaching a maximum $T \approx 300$ K, and cut across the $n=2$ line and all lower lines.  
\item Oda, 1990. \cite{oda1990common}  Peak in the magnetic susceptibility. We omit this data set because it seems have been superceded by Ref. \cite{PhysRevB.49.16000} by the same authors.  
\item Oda, 1990. \cite{oda1990magnetism} Peak in the magnetic susceptibility.  We omit this data set because it duplicates data in Ref. \cite{oda1990common} by the same authors.  
\item Oda, 1991. \cite{oda1991electronic}  Peak in the magnetic susceptibility.  We omit this data set because it seems to have been superceded by Ref. \cite{PhysRevB.49.16000} by the same authors. It does indicate that the peak disappears between $p=0.19$ and $p=0.21$. 
\item Momono, 1996.  \cite{momono1996evidence} Peak in the magnetic susceptibility. We omit this data set because it seems to duplicate data in Ref. \cite{PhysRevB.49.16000}, which was published in the same year by the same authors.  
\item Nakano, 1998. \cite{nakano1998correlation}  Peak in the magnetic susceptibility.  We omit this data set because five out of six data points are the same as Ref. \cite{PhysRevB.49.16000} by the same authors, but divided by $4.3$ in order to map to an energy scale.   
\item Nakano, 1998. \cite{nakano1998correlation}   Magnetic susceptibility.  The temperature reported marks a break from the linear form seen at high temperatures.  We omit this data set because it is the same as the susceptibility data in Ref. \cite{PhysRevB.49.16000} by the same authors, divided by $4.3$ in order to map to an energy scale.   
\item Nakano, 1998. \cite{nakano1998correlation}    Resistivity.  We omit this data set it is roughly a factor of six smaller than the resistivity data in Ref. \cite{PhysRevB.49.16000} by the same authors.   The authors probably divided this data by some number to map to an energy scale, just as they did with the magnetic susceptibility data.  
\item  Hwang, 1994. \cite{PhysRevLett.72.2636} Hall resistance. We omit this data set because the method used to obtain it seems to allow for renormalization of the entire data set by a somewhat arbitrary multiplicative factor.  
\item Batlogg, 1994.   \cite{batlogg1994charge} Hall resistance. We omit this data set because the method used to obtain it seems to allow for renormalization of the entire data set by a somewhat arbitrary multiplicative factor.  The authors are the same as in  Ref. \cite{PhysRevLett.72.2636} and the Hall resistance data looks the same too.  However here the data are limited to $p=0.15$ and higher while the other article includes five additional dopings that are less that $p=0.15$. 
\item Batlogg, 1994. \cite{batlogg1994charge}  Reanalysis of Yoshizaki's magnetic susceptibility data, looking for a shoulder rather than a peak.  We omit this data set because for the two highest doping data points the shoulder is hard to identify and because at the four lower doping data points the physical meaning of the shoulder is not clear, since it remains to be seen whether there is actually a peak at those dopings.  
\item Batlogg, 1994. \cite{batlogg1994charge}  Resistivity - "changes of the high temperature slope curves."  We omit this data set  because it is only at lower dopings and its mathematical and physical meaning, i.e. the precise experimental meaning of "changes of the high temperature slope curves",   is not clear. 
\item Tallon, 1999, analyzes Boebinger, 1996. \cite{tallon1999critical,PhysRevLett.77.5417} Resistivity.  We omit this data set because the method used to obtain it seems to allow for renormalization of the entire data set by an arbitrary multiplicative factor.  
\item Johnston, 1989. \cite{PhysRevLett.62.957} Peak in the magnetic susceptibility.  While this paper is distinguished by being perhaps the first to notice a spin pseudogap, only three data points concern strontium doped LSCO. Of these three points, only one is from data that actually showed a peak in the susceptibility.  We omit the remaining data point.  
\item Takemura, 2000, analyzes Nishikawa, 1994. \cite{takemura2000scaling,nishikawa1994transport} Thermoelectric power, analyzed with a universal scaling method.  We do not plot this data because the temperature reported here lies in the tail of the universal scaling curve and is not associated with any clear feature.  On the other hand the results here are  roughly the same as those produced by \cite{kim2004two}'s thermoelectric power experiment, which used the onset of a linear signal to define their characteristic temperature.  
\item Startseva, 1999. \cite{PhysRevB.59.7184} Optical reflectivity and conductivity. We omit this data because there are only two data points at dopings  separated by only $0.01$ (probably close to the error bars in doping), and because the difference in temperature is very large: $50$ K.
\item Wang, 2006. \cite{PhysRevB.73.024510}    Nernst effect.  The temperature reported here marks a  break from the linear signal seen at higher temperatures.   We omit this paper and several others by the same group because their data has recently been questioned and re-interpreted by Ref. \cite{PhysRevB.97.064502}.
\end{enumerate}
\end{enumerate}

\section{YBCO Data Sets\label{YBCOAppendix}}
 The temperatures gathered here were all realized experimentally rather than by extrapolation from lower temperatures, and clearly identifiable signals  occurred at the reported temperatures, including peaks, kinks, extinction of diffraction peaks, etc.   In the interest of clarity we do not rely on universality arguments, and therefore restrict ourselves to oxygen doped YBCO, and we  keep separate the results of distinct experimental groups and of distinct experimental probes and signatures. 
We do not report any temperatures that are not already reported by the articles we have cited.  In particular, we have stayed out of the business of re-analyzing or fitting data sets from other articles.  The one exception to this rule is our use of the color maps in Ref. \cite{PhysRevLett.93.267001} - from that article we extracted data from certain contours and features that are prominent in the color maps.

 Our survey of pseudogap temperature measurements does not extend to the extensive  literature on anomalies and phase transitions measured using mechanically-oriented observables such as internal friction, sound velocity, lattice constants, thermal expansivity, and the like. \cite{dominec1993ultrasonic,PhysRevLett.85.2376} Nor did we explore the extensive literature on hysteresis in YBCO and associated onset temperatures, or on oxygen movement and ordering.

When the hole doping was not reported, we used Ref. \cite{PhysRevB.73.180505} to map from oxygen content (or, in one case, from $T_c$) to hole doping.

All data sets used in the YBCO figures or enumerated here, with discussion of their particulars and origin, and with a script that produces the figures, are available in the supporting material as a python script.

\begin{enumerate}
\item $n=1$ line:  
\begin{enumerate}
\item Ando, 2004. Resistivity. \cite{PhysRevLett.93.267001}  We reproduce a white contour that is prominent in Ando's data and starts near $p=0.140, \; T=291$ K.  
\end{enumerate}
\item $n=2$ line: In YBCO the $n=2$ line is marked by broken $C_4$, $C_2$, mirror, and inversion symmetries, by fluctuating intra unit cell order which includes time reversal symmetry breaking, by a new contribution to nematic ordering, and by signatures in transport and in  crystal vibrational frequencies.   Several of the symmetry breaking signatures  drop to the $n=4$ line at dopings lower than $ p < 0.085$.

 For this line only we omitted data sets which, compared to their linear regressions, showed a scatter of more than $10$ to $15$ K.
\begin{enumerate}
\item Zhang, 2018. \cite{zhang2018discovery} Muon spin relaxation.  At temperatures below this temperature the spin relaxation reflects the existence of a slowly fluctuating magnetic field consistent with the intra unit cell order seen by neutron scattering, while above this temperature the field is absent.  The entirety of this data is contested by Ref. \cite{sonier2017comment} and a reply to that is contained within Ref. \cite{zhang2018discovery}.  
\item Sato, 2017. \cite{sato2017thermodynamic} Torque magnetometry measurements of the anisotropic susceptibility.  Above this temperature the anisotropy is a linearly increasing function of temperature, while below this temperature it begins increasing as temperature decreases. This signals that at the pseudogap a new   contribution to nematic order is added. 
\item Zhao, 2016. \cite{zhao2017global} Linear and $n=2$ line optical anisotropy.  Below this temperature the $n=2$ line appears, signaling loss of inversion and $C_2$ two-fold rotation symmetries. This is a stronger symmetry breaking than either nematic order or orthorhombic symmetry. 
\item Zhao, 2016, analyzes Lubashevsky, 2014.  \cite{zhao2017global,PhysRevLett.112.147001}  Optical birefringence. The temperature reported here marks the onset of a polarized signal seen at lower temperatures, which signals loss of both mirror and  $C_4$ four-fold rotation  symmetries. 
\item Sidis-Bourges, 2006-2015. \cite{PhysRevLett.96.197001, PhysRevLett.118.097003, PhysRevB.78.020506,PhysRevB.83.104504,mangin2015intra,sidis2007search}   Polarized neutron scattering showing the onset of intra unit cell order, i.e.  time reversal symmetry breaking while retaining lattice translational symmetry.   Thirteen points from six different papers.  The lowest doping data point, at $p=0.08,\, T=170$ K, lies near the $n=4$ line (about $13$ K higher, within the published error bars of $\pm 30$ K), while the other twelve points at higher dopings lie on the $n=2$ line.   We omit this because the data set shows a scatter (compared to its linear regression, and not including the $p=0.08$ data point) of $30$  K, which is above our cutoff.  
\item Daou, 2018/2010,  analyzes Ando, 2004.  \cite{daou2010broken,PhysRevB.97.064502,PhysRevLett.93.267001}  Resistivity.   The temperature reported here marks a  break from the linear signal seen at higher temperatures.   We omit the $p=0.18$ data point on a specially irradiated sample in Ref. \cite{PhysRevB.97.064502}.  
\item Arpaia, 2018. \cite{PhysRevMaterials.2.024804} Thin film on an MgO substrate.  Resistivity.  The temperature reported here marks a transition  to linearity at higher temperatures. 
\item Arpaia, 2018. \cite{PhysRevMaterials.2.024804} Thin film on an SrTiO$_3$ substrate.  Resistivity.  The temperature reported here marks a transition to linearity at higher temperatures. 
\item Alloul, 2010.  \cite{alloul2010superconducting} Resistivity.  The temperature measured here marks a transition to the linear resistivity seen at higher temperatures. We omit the $p=0.169$ data point because the sample has been irradiated to produce more disorder. 
\item Wang, 2017.  \cite{wang2017revisiting} Resistivity. The temperature recorded here marks the transition to quadratic behavior at low temperatures.  
\item Kabanov, 1999.  \cite{PhysRevB.59.1497} Thin films on Mg0 and SrTiO$_3$.  Photoinduced transmission.  The temperature measured here marks the end of a low temperature plateau in $\delta T / T$, where $T$ is the photoinduced transmission signal.  We omit this because the data shows a scatter (compared to its linear regression) of $30$ to $40$  K, which is above our cutoff. 
\item Leridon, 2009. \cite{leridon2009thermodynamic}  First derivative of the magnetic susceptibility with respect to temperature.   At temperatures above this temperature the derivative is a decreasing function of $T$, while at lower temperatures it reverses its behavior and begins decreasing as $T$ is reduced.  We omit this data set because the data shows a scatter (compared to its linear regression) of $30$ to $40$ K, which is above our cutoff.  
\item Shekhter, 2013. \cite{shekhter2013bounding} Resonant ultrasound spectroscopy measuring crystal resonance frequencies. At this temperature a sharp anomaly is seen in the resonance frequency and its slope changes abruptly.  The width of the anomaly is $3$ K, which is far sharper than any other data set on the pseudogap.  

It is worth noting that, unlike the other pseudogap signatures discussed in this article which focus on electronic response, resonant ultrasound spectroscopy is a probe of ionic motion.  As such it belongs  to the extensive cuprate literature on mechanically-oriented observables such as internal friction, sound velocity, lattice constants, thermal expansivity, oxygen movement, thermal history dependence, and the like. \cite{dominec1993ultrasonic,PhysRevLett.85.2376} It is extremely well attested that these observables reveal many distinct anomalies and phase transitions in the temperature range between $T_c$ and room temperature. These features, their dependence on doping, and their qualitative behavior are not yet well understood.  In this connection Ref. \cite{cooper2014pseudogap} has contested Ref. \cite{shekhter2013bounding} 's pseudogap data in its entirety.
\end{enumerate}
\item $n=3$ line: In YBCO the $n=3$ line is marked by time reversal symmetry breaking, onset of a spin resonance,  and a new contribution to nematic order as seen in transport.  
\begin{enumerate}
\item Kapitulnik, 2009.  \cite{kapitulnik2009polar}  This temperature marks the onset of the Kerr effect, signaling time reversal symmetry breaking.  We omit the data point at $p=0.156$ because it has an enormous $87$ K error bar. 
\item Xia, 2008.  \cite{PhysRevLett.100.127002}  Polar Kerr effect.  This temperature marks the onset of the Kerr effect, signalling time reversal symmetry breaking.  They also find hysteresis at these temperatures, up to room temperature, indicating that time reversal symmetry breaking occurs also up to room temperature.   We omit this data set because three out of four data points are repeated in \cite{kapitulnik2009polar}  by the same authors. 
\item Dai, 1999. \cite{dai1999magnetic} Neutron scattering.   The temperatures recorded here mark the onset of a magnetic resonance which is measured by integrating the magnetic structure factor over momentum and frequency. 
\item Cyr-Choiniere, 2015. \cite{PhysRevB.92.224502}  Nematic component of the Nernst effect.  The temperature recorded here marks a transition from steeply decreasing behavior (at low T) to slowly increasing linear behavior (at higher T).  The authors argue that this marks the onset of a new contribution to the nematicity, and that this contribution is distinct from the nematicity at higher dopings which may be associated with charge density waves. 
\item Cyr - Choiniere, 2015, analyzes Ando, 2004.  \cite{PhysRevB.92.224502,PhysRevLett.93.267001}  Nematic component of the resistivity.   They plot the ratio $\rho_a/\rho_b$, where $\rho_a $ and $\rho_b$ are measured along two different axis. The temperature recorded here marks a transition from steeply decreasing behavior (at low T) to slowly increasing linear behavior (at higher T).  The authors argue that this marks the onset of a new contribution to the nematicity, and that this contribution is distinct from the nematicity at higher dopings which may be associated with charge density waves.  
\item Cyr - Choiniere, 2015.  \cite{PhysRevB.92.224502}  Nematic component of the resistivity.   They plot the ratio $\rho_a/\rho_b$, where $\rho_a $ and $\rho_b$ are measured along two different axis. The temperature recorded here marks a transition from a steeply decreasing behavior (at low T)  to slowly increasing behavior (at higher T). The authors argue that this marks the onset of a new contribution to the nematicity, and that this contribution is distinct from the nematicity at higher dopings which may be associated with charge density waves.  
\item Wuyts, 1996. \cite{PhysRevB.53.9418}  Resistivity.  They take the derivative of the resistivity with respect to temperature, find a peak in the derivative, and the peak position is the temperature recorded here.  They originally multiplied by two and we remove that factor.  
\end{enumerate}
\item $n=4$ line: In YBCO the extent of the $n=4$ line, from $p=0.053$ to $p=0.142$, is attested to by three transport data sets.
\begin{enumerate}
\item Arpaia, 2018. \cite{PhysRevMaterials.2.024804}  Thin film on an MgO substrate.  Resistivity.  The temperature reported here marks a transition from quadratic at lower temperatures.  Data from $p=0.053$ to $p=0.142$. 
\item Arpaia, 2018. \cite{PhysRevMaterials.2.024804} Thin film on an SrTiO$_3$ substrate.  Resistivity.  The temperature reported here marks a transition from quadratic at lower temperatures. Data from $p=0.067$ to $p=0.140$. 
\item LeBouef, 2011, analyzes Segawa, 2004.  \cite{PhysRevB.83.054506,PhysRevB.69.104521}   Hall resistance. This temperature records the point where the second derivative of the Hall resistance changes sign. Data from $p=0.055$ to $p=0.119$.  
\end{enumerate}
At low dopings from $p=0.052$ to $p=0.082$ the $n=4$ line is augmented by  intra unit cell order,  spontaneous magnetic fields, and new nematic order. At higher dopings similar signals are seen on the $n=2$ line.
\begin{enumerate}
\item Haug, 2010.  \cite{haug2010neutron}  Neutron scattering.   The temperature reported here marks onset of anisotropy in a neutron scattering triple-axis experiment. This is a sign of nematic order, and is called an electronic liquid crystal. 
\item Baledent, 2011. \cite{PhysRevB.83.104504} Polarized neutron scattering showing the onset of intra unit cell order, i.e.  time reversal symmetry breaking while retaining lattice translational symmetry.  The lowest doping data point, at $p=0.08,\, T=170$ K, lies near the $n=4$ line (about $13$ K higher, within the published error bars of $\pm 30$ K).  
\item Sonier, 2001.   \cite{sonier2001anomalous} Muon spin relaxation.  This is the extinction temperature of a signal that indicates the presence of small spontaneous magnetic fields.  Two data points.  The first lies on the $n=4$ line. We omit the second data point, which lies about $35$ K below the $n=4$ line, well outside the experimental error bars of $\pm 10$ K, and close to the three dimensional charge density waves \cite{laliberte2018high}.  
\end{enumerate}
\item Other Data Sets:
\begin{enumerate}
\item Arpaia, 2018.  \cite{PhysRevMaterials.2.024804} Thin film on an MgO substrate.  Resistivity.  The temperature reported here marks a transition to quadratic  at higher temperatures. We omit this data from the linear regressions because it clearly has two parts, one at lower doping which decreases very steeply until it hits the superconducting doping, and a second part which follows the superconducting dome.
\item Arpaia, 2018.  \cite{PhysRevMaterials.2.024804} Thin film on an SrTiO$_3$ substrate.  Resistivity.  The temperature reported here marks a transition to quadratic at higher temperatures. 
\end{enumerate}
\item Superconducting $T_c$, Neel temperature, and charge density waves:
\begin{enumerate}
\item Coneri, 2010. \cite{PhysRevB.81.104507}    Neel temperature measured with muon spin rotation. 
\item Coneri, 2010.  \cite{PhysRevB.81.104507} $T_c$ measured with muon spin rotation. 
\item Liang, 2006.   \cite{PhysRevB.73.180505} $T_c$. 
\item Laliberte, 2018.   \cite{laliberte2018high} Sound velocity measurements of $3-$D charge density wave order. We omit the $T=0$ data points at either side of the dome.
\end{enumerate}
\item Omitted Data Sets:
\begin{enumerate}
\item Hinkov, 2008. \cite{hinkov2008electronic} Neutron scattering. Polarized neutron scattering showing the onset of intra unit cell order, i.e.  time reversal symmetry breaking while retaining lattice translational symmetry.  We omit this data point because it seems to have been revised from $T=150$ K to $T = 170$ K in Ref. \cite{PhysRevB.83.104504} by the same authors..
\item Cyr-Choiniere, 2018 and Daou, 2010. \cite{PhysRevB.97.064502,daou2010broken}  Nernst effect.   The temperature reported here marks a  break from the linear signal seen at higher temperatures.   In \cite{daou2010broken} it is shown  that this temperature is the point where anisotropy in the Nernst coefficient is extinguished,  at $p=0.12, \, 0.13, \, 0.15, \, 0.18$.  The authors argue that this anisotropy is caused by rotational symmetry breaking in the copper oxide planes as opposed to the oxygen chains.    Ref. \cite{PhysRevB.92.224502} by the same authors, five years later, re-evaluates the data, and says that the Nernst anisotropy seen in Ref. \cite{daou2010broken} "is more likely to be caused by CDW modulations" instead of the pseudogap.   In other words, although they saw the beginning of a slight rise in the Nernst anisotropy at the pseudogap temperature, the real rise doesn't occur until near the lower temperatures where CDW order is observed using X-ray diffraction.  Ref. \cite{PhysRevB.92.224502} is an effort to sort out where the new nematicity begins. We omit this data set (but not Ref. \cite{PhysRevB.92.224502}) because it was later revised and reinterpreted by the same authors.  We also note  that  the data shows a scatter (compared to its linear regression) of $15$ to $20$ K, which is above our cutoff for the $n=2$ line.  This data set cuts across the $n=1$ and $n=2$ lines and is rather flat. 
\item Goto, 1996.   \cite{goto1996carrier}  Nuclear magnetic resonance.  The temperature reported here marks a peak in the signal.  This data set is very flat and cuts through the $n=2$ and $n=3$ lines. It roughly coincides with X-ray scattering data on short-range charge density wave order from Ref. \cite{PhysRevB.90.054513}.  
\item Cooper, 1996. \cite{cooper1996some}  Several pseudogap temperature data sets are reported in Figure 30.  This data is omitted because we don't understand where it came from and therefore can not verify whether it meets our selection criteria, and also because it seems that the thermoelectric power data in Figure 30 (which runs up to $600$ K) was derived from data in Figure 28b, which has a temperature cutoff of $300$ K.
\end{enumerate}
\end{enumerate}


\begin{acknowledgments}
We gratefully acknowledge  stimulating discussions with K. S. Kim, J. Zaanen,  A. Ho, S. Hayden, M. Sulangi, K. Schalm,  V. Cheianov, T. Ziman, I. Vishik, J. Saunders, L. Levitin, J. Koelzer, P. Hasnip, M. Ma, P. Abbamonte, A. Kim, P. Coleman, S. Hayden, and especially A. Petrovic.  We thank the Instituut Lorentz in Leiden for hospitality, and the Hubbard Consortium for facilitating discussions.    We acknowledge  support from EPSRC grant EP/M011038/1. 
  \end{acknowledgments}



\bibliography{vincent}

\end{document}